\documentclass[prb,aps,amsmath,preprint]{revtex4}

\usepackage{graphicx}
\usepackage[version=3]{mhchem}
\usepackage{dcolumn}
\usepackage{multirow}
\usepackage{bm}
\usepackage{subfigure}

\begin{document}
\title{Computational study of structural, electronic and optical properties of crystalline NH$_4$N$_3$}
\author{N. Yedukondalu, Vikas D. Ghule, and G. Vaitheeswaran$^*$ }

\affiliation{Advanced Centre of Research in High Energy Materials (ACRHEM),
University of Hyderabad, Prof. C. R. Rao Road, Gachibowli, Andhra Pradesh, Hyderabad- 500 046, India.}
\date{\today}

\begin{abstract}
A systematic computational study on the structural, electronic, bonding, and optical properties of orthorhombic ammonium azide (NH$_4$N$_3$) has been performed using planewave pseudopotential (PW-PP) method based on density functional theory (DFT). Semiempirical dispersion correction schemes have been used to account for non-bonded interactions in molecular crystalline NH$_4$N$_3$. The ground state lattice parameters and fractional co-ordinates obtained using the dispersion correction schemes are in excellent agreement with experimental results. We calculated the single crystal elastic constants of NH$_4$N$_3$ and its sensitivity is interpreted through the observed ordering of the elastic constants (C$_{33}$ $\textgreater$ C$_{11}$ $\textgreater$ C$_{22}$). The electronic structure and optical properties were calculated using full potential linearized augmented plane wave (FP-LAPW) approach with recently developed functional of Tran-Blaha modified Becke-Johnson (TB-mBJ) potential. The TB-mBJ electronic structure shows that NH$_4$N$_3$ is a direct band gap insulator with a band gap of 5.08 eV, while the calculated band gap with standard generalized gradient approximation is found to be 4.10 eV. The optical anisotropy is analyzed through the calculated optical constants namely dielectric function and refractive index along three different crystallographic axes. The absorption spectra reveal that NH$_4$N$_3$ is sensitive to ultraviolet (UV) light. Further, we also analyzed the detonation characteristics of the NH$_4$N$_3$ using the reported heat of formation and calculated density. NH$_4$N$_3$ is found to have a detonation velocity of 6.45 km/s and a detonation pressure about 15.16 GPa computed by Kamlet-Jacobs empirical equations. \\
\end{abstract}

\maketitle
\section {Introduction}
 NH$_4$N$_3$ belongs to the class of inorganic azides, which are an interesting group of materials with a wide range of applications as explosives and photographic materials. They served as model systems for studying the fast reactions in crystalline solids with complex chemical bonding.\cite{fair, yoffe} Alkali metal azides in particular sodium azide can be used to synthesis polymeric nitrogen under extreme conditions, which is a green high energy density material (HEDM)\cite{hemley, eremets}, and it can be used as rocket fuel or propellant.\cite{trojen} Among the inorganic azides, NH$_4$N$_3$ is of special interest due to the strong hydrogen bonding features between the ammonium cation and the negatively charged azide anion, which can be considered as model hydrogen-bonded system.\cite{medvedev, palasyuk, prince} NH$_4$N$_3$, first obtained by Curtius\cite{curtius} is a highly potential azide for its ability to detonate powerfully with a very low sensitivity. NH$_4$N$_3$ is thermally unstable inorganic azide, which is highly volatile and undergoes molecular sublimation. Recently, Eslami \emph{et al}\cite{eslami} reported the microencapsulation techniques to stabilize volatile   NH$_4$N$_3$, while Ng \emph{et al}\cite{ng} made an early comprehensive experimental analysis of NH$_4$N$_3$ based on the sublimation kinetics over the temperature range 360-389 K. The reported differential scanning calorimetry thermogram results confirms the endothermic (73.4 kJ/mol) nature of NH$_4$N$_3$, and its vapor readily dissociates into ammonia and hydrazoic acid.\cite{ng,evers}

\par Materials that contains hydrogen and nitrogen alone are considered to be promising candidates for HEDMs as their decomposition reactions gives environmental friendly gases such as nitrogen and hydrogen. The high volatile nature, thermal instability, and impact sensitivity of NH$_4$N$_3$ are the main disadvantages for its usage in energetic material formulations. However, NH$_4$N$_3$ is capable of considerable brisance\cite{yakovleva} owing to the rapid evolution of gaseous products. Apart from its usage as a weak explosive, its potential as a gas generator has received much attention in recent years. NH$_4$N$_3$ has been used to inflate safety cushions in automobiles with a suitable oxidizer\cite{masaaki} and also used as a solid propellant in photochemical micro-rockets for altitude control.\cite{maycock} NH$_4$N$_3$ is a derivative of nitric-hydro-acids, which is an important component of rocket fuels.\cite{sarner}  Moreover, recent experiments\cite{medvedev, palasyuk} reveal that NH$_4$N$_3$ undergoes a polymorphic phase transition about 3 GPa due to modification in the strength of hydrogen bonding and also the high pressure phase is thermodynamically stable up to 55 GPa. While ab initio studies predict the high pressure phase of NH$_4$N$_3$ to nonmolecular hydro-nitrogen solid at 36 GPa, which is a HEDM with wide range of applications.\cite{hu}         

\par Over the past several decades, a number of studies have been devoted to the structural properties and decomposition mechanism of NH$_4$N$_3$.\cite{ng, kurgangalina, finch} However, many fundamental aspects of NH$_4$N$_3$ are still not well understood because of its complex chemical behavior. The detailed knowledge about the crystal structure of azides is very important in order to understand the stability and decomposition. One of the primary objectives of this study is to understand ground state electronic structure and optical properties of NH$_4$N$_3$, which are not investigated so far at the ab initio level. 

\par The rest of article is organized as follows: in section II, we briefly describe the computational details. Results and discussion concerning structural and elastic properties of NH$_4$N$_3$ are presented in section III-A, section III-B and III-C contains the details of electronic structure, bonding and optical properties of NH$_4$N$_3$. In section III-D, we discuss the detonation characteristics of NH$_4$N$_3$, and finally, section IV summarizes the conclusions.

\section{Computational details}
First-principles calculations were performed by using two distinct approaches known as PW-PP and FP-LAPW methods. Structural and elastic properties were obtained using PW-PP approach as implemented in Cambridge Series of Total Energy Package\cite{Payne,Milman,Segall} based on DFT.\cite{Hohenberg, Kohn} We used Vanderbilt-type\cite{Vanderbilt} ultrasoft pseudopotentials for electron-ion interactions. The local density approximation (LDA)\cite{Ceperley,Perdew} and generalized gradient approximation (GGA)\cite{Burke} were used to treat electron-electron interactions in NH$_4$N$_3$. The standard LDA (CA-PZ) and GGA (PBE) functionals are inadequate to predict the long range interactions in molecular crystalline solids.\cite{Santra, Lu, Dion} However, the non covalent interactions such as hydrogen bonding and van der Waals interactions play a key role in determining the physical and chemical properties of molecular solids. Hence, the semiempirical approaches have been developed in order to account for the long range interactions in the molecular solids and they were incorporated through standard DFT description. In the present study, we have used two dispersion schemes. First, G06 scheme by Grimme,\cite{Grimme} which is an empirical correction to DFT taking into account of the dispersive interactions based on damped and atomic pairwise potentials of the form C$_6$.R$^{-6}$. Second, TS scheme recently developed by Tkatchenko and Scheffler\cite{Tkatchenko} based on the summation of interatomic C$_6$ coefficients derived from the electron density of molecule or solid and accurate data for the free atoms. The C$_6$ coefficients used in the calculations, for N and H, atoms are 12.75, 1.451 in G06 and 14.46, 3.884 eV.$\AA^6$ in TS schemes, respectively. We systematically studied the effect of semi empirical dispersion correction schemes on the structural properties of the NH$_4$N$_3$ along with standard DFT functionals, and discuss the same in detail in the following section. The Broyden-Fletcher-Goldfarb-Shanno (BFGS) minimization scheme\cite{Almlof} has been used for structural relaxation. The plane wave basis orbitals used in the calculations are 2$s^2$, 2$p^3$ for N and 1$s^1$ for H. The convergence criteria for structural optimization was set to fine quality with a kinetic energy cutoff of 580 eV and k-mesh 4x6x4 according to the Monkhorst-Pack grid scheme.\cite{Monkhorst} The self-consistent energy convergence was set to 1.0$\times$10$^{-5}$ eV/atom. The convergence criterion for the maximal force between atoms was 0.03 eV/A. The maximum displacement and stress were set to be 1.0$\times$10$^{-2}\AA$ and 0.05 GPa, respectively.

\par It is well known that the standard DFT functionals LDA and GGA usually underestimate the energy-band gap about 50\% when compared to experiments. The LDA and GGA functionals suffer from artificial electron self-interaction and also lack the derivative discontinuities of the exchange-correlation potential with respect to occupation number.\cite{Nieminen} The calculation of band gap involves only exchange energy\cite{Sham} and it was determined very accurately by the TB-mBJ functional\cite{peter} when compared to standard DFT functionals. Hence, this semi local functional provides much improved band gaps for semiconductors and insulators. Therefore, the electronic structure and optical properties of NH$_4$N$_3$ were calculated by using this TB-mBJ functional as implemented in WIEN2k package.\cite{blaha} To achieve energy eigen values convergence, wave functions in the interstitial region were expanded in plane waves with cut-off K$_{max}$ = 4/R$_{MT}$, where R$_{MT}$ is the smallest atomic sphere radius and K$_{max}$ denotes the magnitude of largest k vector in plane wave expansion, while the charge density was Fourier expanded upto G$_{max}$ = 20. The muffin-tin radii were assumed to be 0.7a$_0$, 1.1a$_0$ for H and N atoms, respectively, where a$_0$ is the Bohr radius. Self consistency is obtained using 54 k-points in the irreducible Brillouin zone (IBZ). 

\section{Results and discussion}
\subsection{Structural and elastic properties}
NH$_4$N$_3$ crystallizes in orthorhombic structure with space group $Pmna$ with a = 8.937$\AA$, b = 3.807$\AA$, c = 8.664$\AA$, and Z = 4 at ambient conditions.\cite{amorim} In order to obtain the equilibrium crystal structure, we have performed full structural optimization of NH$_4$N$_3$ including lattice parameters and internal coordinates using standard LDA and GGA functionals. We find a difference between calculated and experimental volume of -9.9$\%$ with LDA and +3.9$\%$ with GGA and thus, our GGA volume is closer to experiment than LDA volume. This discrepancy between theory and experiment is due to standard exchange-correlation potentials used in the calculations that do not capture the nature of non-bonded interactions such as hydrogen bonding and van der Waals interactions in molecular solids.\cite{Santra, Lu, Dion}  Therefore, we have performed structural relaxation with the semiempirical approaches known as G06 and TS schemes to include the non-bonded interactions in our calculations. The volume is underestimated by 3.5$\%$ using G06, this error is approximately the same as the standard GGA value with opposite sign. However, the TS scheme provides much better volume (293.99$\AA^3$) 0.3$\%$ lesser than the experimental volume (294.78$\AA^3$), where the deviations in lattice constants a, b, c are +0.9$\%$, -0.02$\%$, and -1.2$\%$, respectively. The optimized equilibrium crystal structure is in Fig. 1. The calculated lattice parameters, volume and fractional co-ordinates with standard DFT functionals and semiempirical schemes are presented in Tables I and II, respectively. We have obtained equilibrium bulk modulus $B_0$ = 26.34 GPa, and its pressure derivative, $B^\prime_0$ = 3.6, by fitting pressure-volume data to Murnaghan equation of state.\cite{murnaghan} The bulk modulus value of NH$_4$N$_3$ lies between alkali metal azides (AMAs) and  heavy metal azides (HMAs); the reported experimental bulk moduli for AMAs, LiN$_3$ (19.1 GPa),\cite{medevdev} NaN$_3$ (16.3 GPa),\cite{weir} KN$_3$ (18.6 GPa),\cite{cheng} and CsN$_3$ (18 GPa)\cite{dongbin} and for HMAs, AgN$_3$ (39 GPa),\cite{hou} and $\alpha$-Pb(N$_3)_2$ (41 GPa),\cite{weir} which indicates that NH$_4$N$_3$ is a softer material than HMAs and harder than AMAs. The bulk modulus of NaN$_3$ and $\alpha$-Pb(N$_3)_2$ is calculated from their compresibilities.\cite{weir} Overall, the present study reveals that TS scheme works better for molecular crystalline NH$_4$N$_3$. Hence, we have used this equilibrium volume to calculate the elastic properties of NH$_4$N$_3$.      

\begin{figure}[h]
\centering
\includegraphics[height = 3.75in, width=6.5in]{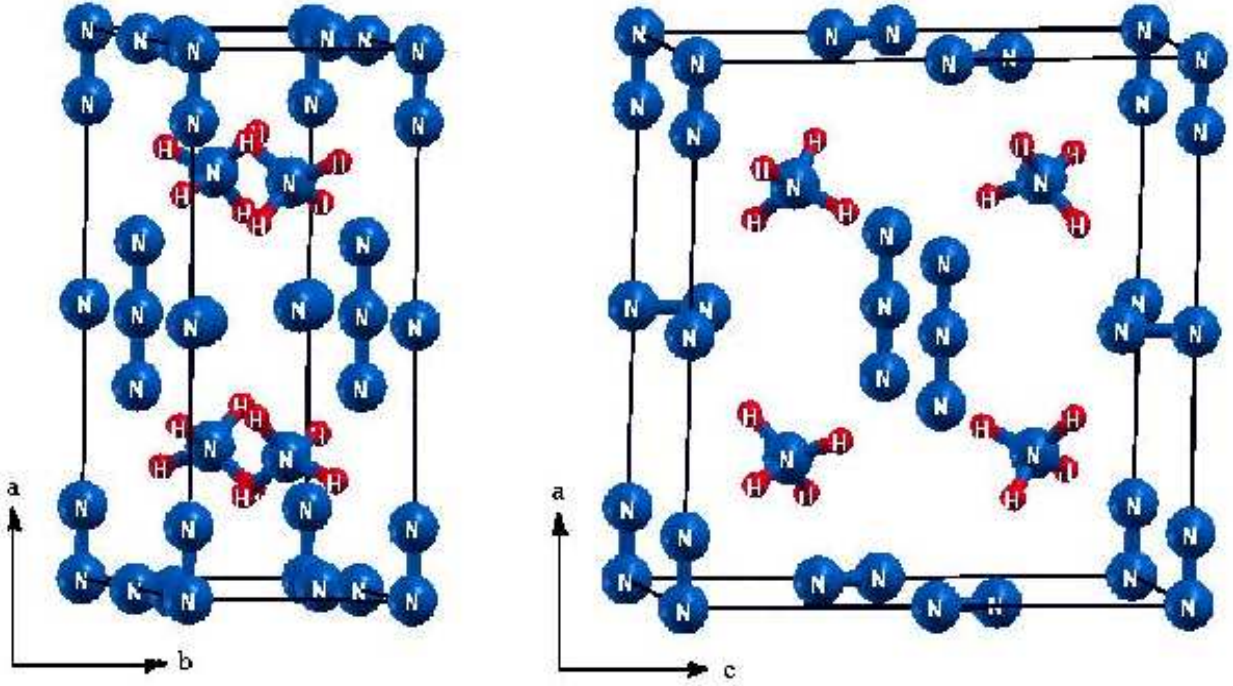}
\caption{(colour online) Crystal structure of ammonium azide along i) a-b and ii) a-c-axis (Blue and red colours represent N and H atoms, respectively). }
\end{figure}
 
\begin{table}[h]
\caption{Calculated Ground State Lattice parameters (a, b, c, in $\AA$) and Volume (V, in $\AA^3$) of NH$_4$N$_3$ Using Standard DFT Functionals LDA (CA-PZ), GGA (PBE) and Semiempirical Schemes PBE+G06, PBE+TS Implemented Thorugh GGA Functional along with Experimental Data.}
\begin{ruledtabular}
\begin{tabular}{cccccc}
Parameter  &      CA-PZ       &   PBE        &    PBE+G06  &  PBE+TS   &  experiment.\cite{amorim}  \\ \hline
a   &     8.743        &  9.094       &    8.904    &  9.019    &  8.937  \\
b   &     3.631        &  3.904       &    3.722    &  3.806    &  3.807  \\
c   &     8.362        &  8.627       &    8.585    &  8.563    &  8.664  \\
V   &     265.52       &  306.36      &    284.57   &  293.99   &  294.78 \\
\end{tabular}
\end{ruledtabular}
\end{table}

\begin{table}[h]
\caption{Calculated Fractional Coordinates of NH$_4$N$_3$ Using Standard DFT Functionals and Semiempirical Schemes along with Experiment.\cite{amorim}}
\begin{ruledtabular}
\resizebox{0.8\textwidth}{!}{
\begin{tabular}{ccccc}
Atom           &    Method        &     x      &      y    &    z        \\  \hline
               &   CA-PZ          &  0.0000    &   0.0000  &  0.0000     \\
               &    PBE           &  0.0000    &   0.0000  &  0.0000     \\
N(M) of N$_3$  &   PBE+G06        &  0.0000    &   0.0000  &  0.0000     \\
               &   PBE+TS         &  0.0000    &   0.0000  &  0.0000     \\
               &   exptl          &  0.0000    &   0.0000  &  0.0000     \\

               &    CA-PZ         &  0.5000    &   0.0000  &  0.0000     \\
               &     PBE          &  0.5000    &   0.0000  &  0.0000     \\
N(M) of N$_3$  &   PBE+G06        &  0.5000    &   0.0000  &  0.0000     \\
               &   PBE+TS         &  0.5000    &   0.0000  &  0.0000     \\
               &   exptl          &  0.5000    &   0.0000  &  0.0000     \\  
          
               &    CA-PZ         &  0.1382    &   0.0000  &  0.0000     \\
               &     PBE          &  0.1341    &   0.0000  &  0.0000     \\
N(T) of N$_3$  &   PBE+G06        &  0.1366    &   0.0000  &  0.0000     \\
               &   PBE+TS         &  0.1348    &   0.0000  &  0.0000     \\
               &    exptl.        &  0.1315    &   0.0000  &  0.0000     \\          
     
               &    CA-PZ         &  0.5000    &   0.1253  &  0.1338     \\
               &     PBE          &  0.5000    &   0.1435  &  0.1249     \\
N(T) of N$_3$  &    PBE+G06       &  0.5000    &   0.1159  &  0.1324     \\
               &    PBE+TS        &  0.5000    &   0.1424  &  0.1270     \\
               &     exptl.       &  0.5000    &   0.1103  &  0.1320     \\          

               &    CA-PZ         &  0.2500    &   0.5516  &  0.2500     \\
               &     PBE          &  0.2500    &   0.5537  &  0.2500     \\
N of NH$_4$    &   PBE+G06        &  0.2500    &   0.5474  &  0.2500     \\
               &   PBE+TS         &  0.2500    &   0.5556  &  0.2500     \\
               &    exptl         &  0.2500    &   0.5524  &  0.2500     \\
          
               &     CA-PZ        &  0.2896    &   0.7206  &  0.3450     \\
               &     PBE          &  0.2914    &   0.7104  &  0.3388     \\
H(1)           &   PBE+G06        &  0.2902    &   0.7108  &  0.3408     \\
               &    PBE+TS        &  0.2911    &   0.7152  &  0.3403     \\
               &     exptl        &  0.2900    &   0.7310  &  0.3330     \\          
          
               &    CA-PZ         &  0.3409    &   0.3842  &  0.2079     \\
               &     PBE          &  0.3349    &   0.3989  &  0.2058     \\
H(2)           &   PBE+G06        &  0.3375    &   0.3850  &  0.2073     \\
               &   PBE+TS         &  0.3359    &   0.3977  &  0.2059     \\
               &    exptl         &  0.3230    &   0.3880  &  0.1920     \\                
                
\end{tabular}}
\end{ruledtabular}
\end{table}

\par The elastic constants are fundamental parameters for crystalline solids, which describe stiffness of the solid against externally applied strains. The elastic constants were calculated using volume-conserving strains technique \cite{Mehl} with the PBE+TS scheme. A complete asymmetric crystal behavior can be described by 21 independent elastic constants, due to orthorhombic symmetry of the NH$_4$N$_3$ crystal, it has nine independent elastic constants, namely, C$_{11}$, C$_{22}$, C$_{33}$, C$_{44}$, C$_{55}$, C$_{66}$, C$_{12}$, C$_{13}$, and C$_{23}$. The calculated elastic constants are presented in Table III. The calculated elastic (stiffness) constants are positive and obey the Born's mechanical stability criteria\cite{Born},  indicates that NH$_4$N$_3$ is mechanically stable at ambient pressure. Eckhardt et al\cite{haycraft, stevens, eckhardt} have shown the qualitative association of stiffness constants, C$_{11}$, C$_{22}$, and C$_{33}$ of crystal with physical phenomena including cleavage planes, patterns in crystal growth, and molecular packing. The strength of interactions between the molecules comprising a molecular crystal has a measurable effect on the macroscopic properties of the solid. Detonation of energetic material can be considered to be a collective property of the material and is highly dependent upon intermolecular interactions, molecular arrangements, and molecular composition. These properties can often be correlated to strength of the lattice interactions through elastic constants. Since, NH$_4$N$_3$ possesses orthorhombic crystal symmetry, the stiffness constants C$_{11}$, C$_{22}$ and C$_{33}$ can be directly related to the crystallographic a, b, and c axes, respectively. The observed ordering of stiffness constants in NH$_4$N$_3$ is, C$_{11}$ $\approx$ C$_{33}$ $\textgreater$ C$_{22}$. From the present results, C$_{22}$ is found to have the weakest stiffness constant, which represents a relative weakness of lattice interactions along the crystallographic b-axis. The intermolecular interactions are relatively stronger along a and c axes due to the orientation of nitrogen atoms of azide ion in these crystallographic directions (see Fig. 1ii). Further, this can be supported by the higher values of stiffness constants C$_{11}$ and C$_{33}$, when compared to  the stiffness constant C$_{22}$. This results reveal that NH$_4$N$_3$ is sensitive to impact along the b-crystallographic axis. However, C$_{44}$, C$_{55}$, and C$_{66}$ indicate the shear elasticity applied to the two dimensional rectangular lattice in the (100), (010), and (001) planes. From our calculations, C$_{55}$ is found to be relatively small compared to C$_{44}$ and C$_{66}$, which is an indication of the soft shear transformation along (010) plane.       

\par We also derived the polycrystalline properties of NH$_4$N$_3$ such as bulk, shear and Young moduli from the calculated elastic constants using Reuss approximation.\cite{reuss} The bulk modulus (B) measures resistance of a material against volume change under hydrostatic pressure, which indicates the average bond strength of material. The calculated bulk modulus and compressibility are found to be 26.8 GPa and 0.037 GPa$^{-1}$ for NH$_4$N$_3$, respectively. The reported\cite{weir} compressibilities for different inorganic azides are, 0.0615 GPa$^{-1}$ for NaN$_3$, 0.0541 GPa$^{-1}$ for KN$_3$, 0.0463 GPa$^{-1}$ for TlN$_3$, 0.0256 GPa$^{-1}$ for AgN$_3$\cite{hou} and 0.0244 GPa$^{-1}$ for $\alpha$-Pb(N$_3$)$_2$. The calculated results reveal that NH$_4$N$_3$ shows compressibility between AMAs and HMAs. The bulk modulus calculated from the elastic constants is in excellent agreement with the obtained equilibrium bulk modulus from the equation of state. This might be an evidence of reliability and accuracy of our calculated elastic constants for NH$_4$N$_3$. The shear modulus (G) represents the resistance to shape change caused by shearing force, which indicates the resistance to change in the bond angle and it is found to be 12.1 GPa. The Young modulus (E) of a material is defined as the ratio of the linear stress to linear strain, which tells about the stiffness of material. The calculated value 31.6 GPa of NH$_4$N$_3$ reveals that the material is stiffer. In general, Poisson's ratio of a material quantifies the stability of crystal against shear strain. If $\sigma$ is 0.5, no volume change occurs, whereas lower than 0.5 means that large volume change associated with elastic deformation.\cite{ravindran} According to our calculation, the $\sigma$ value is 0.30, which shows that a considerable volume change can be associated with the deformation in NH$_4$N$_3$. According to Pugh's criterion,\cite{pugh} the B/G ratio less than (greater than) critical value 1.75 indicates the brittle (ductile) nature of materials. In the present case, it is found to be 2.2, which implies that NH$_4$N$_3$ is a ductile material. This is also confirmed by the Cauchy's pressure (C$_{12}$-C$_{44}$), the negative and positive values of Cauchy's pressure indicate the brittle and ductile nature of materials, respectively.\cite{pettifor} A positive value of Cauchy's pressure indicating that NH$_4$N$_3$ has ductile nature, as commonly expected for ionic insulators. Further, we also estimated the Debye temperature $\Theta_D$, which is a fundamental quantity that determines the thermal characteristics of material. A high value of $\Theta_D$ implies higher thermal conductivity. At low temperatures $\Theta_D$, can be estimated from the average wave velocities ($v_m$) of longitudinal ($v_l$) and transverse ($v_t$) modes as mentioned in refs. \onlinecite{ravindran, anderson}. The calculated values of $v_l$, $v_t$ and $v_m$ are 5.62, 2.99, and 3.34 km/s respectively, which yields a $\Theta_D$ of 474.5 K (see Table III). This is the first qualitative prediction of elastic properties of NH$_4$N$_3$ that still awaits experimental confirmation. 

\begin{table}[h]
\caption{Calculated Elastic Constants C$_{ij}$ (in GPa), Bulk modulus (in GPa), Shear modulus (in GPa), Density (in g/cc), Experimental Density in Parenthesis and Debye Temperature (in K) of NH$_4$N$_3$ Using PBE+TS Scheme.}
\begin{ruledtabular}
\begin{tabular}{ccccccccccccc}
 C$_{11}$ &  C$_{22}$ & C$_{33}$ &  C$_{44}$  &  C$_{55}$  &  C$_{66}$  & C$_{12}$  &   C$_{13}$ & C$_{23}$  &   B    &  G    & $\rho$ & $\Theta_D$ \\ \hline
   45.7   &    35.6   &  46.3    &   16.1     &   10.6     &   13.3     &  17.6     &     19.1   &   21.9    &  26.8  & 12.1 &  1.357(1.36)\cite{williams}  &   474.5
\end{tabular}
\end{ruledtabular}
\end{table}

\subsection{Electronic structure and chemical bonding }
Electronic structure of inorganic azides have been investigated by X-ray electron spectroscopy\cite{colton} to understand the chemical bonding. This study reveals that the ionic character intensifies in AMAs as follows, LiN$_3$ $\textless$ NaN$_3$ $\textless$ KN$_3$ $\textless$ RbN$_3$ $\textless$ CsN$_3$; this is in very good agreement with recently reported\cite{dongbin} trend for the AMAs, while the same in HMAs, (AgN$_3$, CuN$_3$, Cu(N$_3$)$_2$, Hg(N$_3$)$_2$, Hg$_2$(N$_3$)$_2$) $\textless$  (TlN$_3$, $\alpha$-Pb(N$_3$)$_2$) $\textless$ Cd(N$_3$)$_2$, which indicates that AMAs are more ionic than HMAs, implying that AMAs are relatively stable than HMAs. In order to understand the relative stability of NH$_4$N$_3$ when compared to the mentioned inorganic azides, it is necessary to know the electronic structure and chemical bonding in NH$_4$N$_3$. The electronic structure determines the fundamental physical and chemical properties such as  initiation, decomposition, and detonation of a energetic material. Electronic band gap is an important property for energetic materials, which can be used to predict the relative stability and sensitivity,\cite{zhu1} but the standard DFT functionals always underestimate the band gap approximately 50\% when compared to experiments. Hence, we need more accurate methods such as GW approximation\cite{hedin1,hedin2} in order to get a reliable value for energy gap. Recent studies\cite{peter, koller1, koller2, dixit, singh} show that the TB-mBJ functional provides fairly accurate energy gaps for semiconductors and insulators, and also this method is computationally less expensive when compared to the former method. Therefore, we have used this TB-mBJ functional to calculate electronic structure and optical properties of NH$_4$N$_3$ at ambient pressure. Previous theoretical studies\cite{zhu1,zhu2,zhu3,zhu4} revealed the relationship between band gap and impact sensitivity for the metal azides and CHNO based energetic materials within the framework of periodic DFT calculations. Further, Kuklja et al\cite{kuklja1, kuklja2} investigated the excitonic mechanism of detonation initiation in explosives and clarified that pressure inside the impact wave front reduces the band gap. It can be expected that the smaller the band gap, the easier the electron transfers from valence band to conduction band; thus, the energetic system becomes more and more sensitive to external stimuli such as light, heat, friction, and impact. The calculated band structure of NH$_4$N$_3$ along high symmetry directions is as follows, $\Gamma$-point at 5.08 eV; Z, T, U, R-points at 5.29 eV; Y-point at 5.21 eV and S, X-points at 5.24 eV in the first Brillouin zone as shown in Fig. 2. The top of valence band and bottom of the conduction band occur along $\Gamma$-$\Gamma$ direction, indicating that this material is a direct band gap insulator with a minimum separation of 5.08 eV. The corresponding band gap value using GGA-PBE functional is 4.10 eV. Experimental band gap value when available could be compared to our theoretical band gap. \\

\begin{figure}[h]
\centering
\includegraphics[height = 6.0in, width=5.0in]{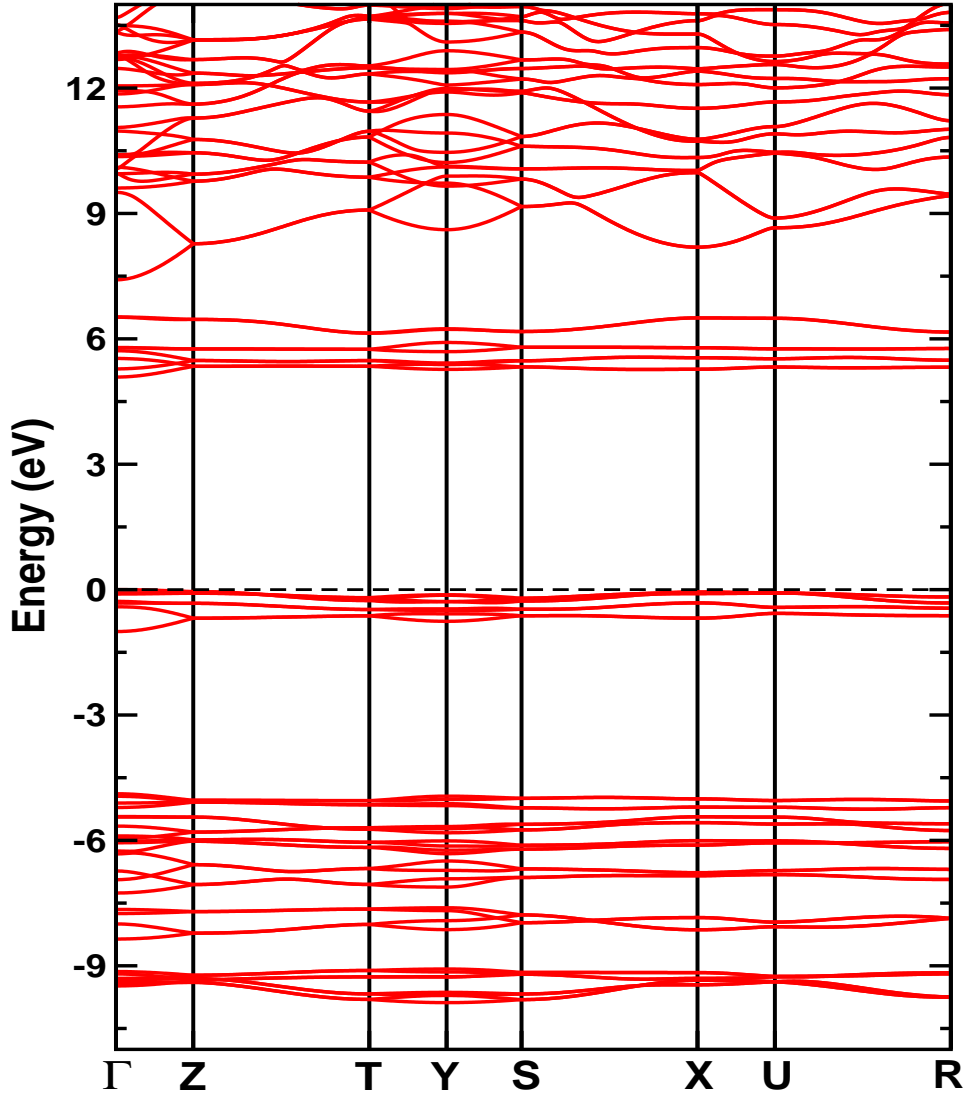}  
\caption{(Color online) The calculated band structure of NH$_4$N$_3$ along high symmetry directions in the Brillouin zone $\Gamma$ (0.0, 0.0, 0.0) $\rightarrow$ Z (0.0, 0.0, 0.5)$\rightarrow$ T (0.0 0.5, 0.5) $\rightarrow$ Y (0.0 0.5 0.0) $\rightarrow$ S (0.5, 0.5, 0.0) $\rightarrow$ X (0.5, 0.0, 0.0) $\rightarrow$ U (0.5, 0.0, 0.5) $\rightarrow$ R (0.5, 0.5, 0.5) using the TB-mBJ functional at the experimental crystal structure}
\end{figure}

\begin{figure}[h]
\centering
\includegraphics[height = 4.5in, width=5.5in]{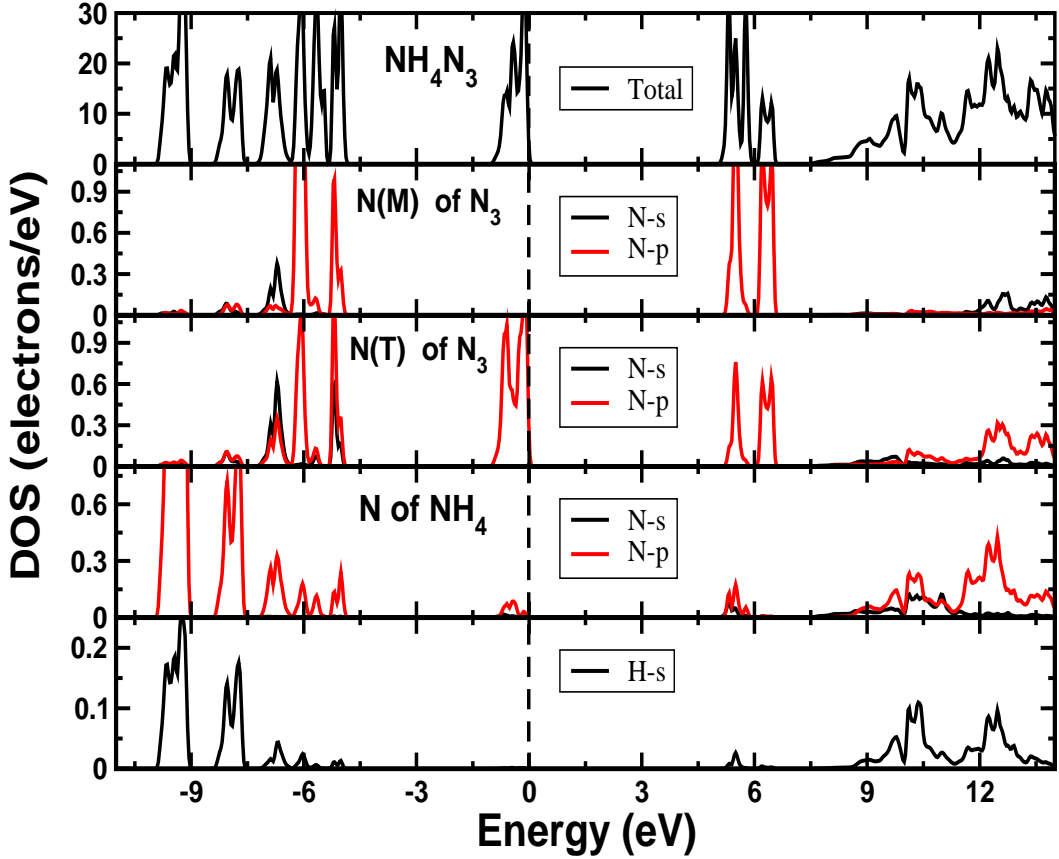}
\caption{(Color online) The calculated total and PDOS of NH$_4$N$_3$ using the TB-mBJ functional at the experimental crystal structure. N(T) and N(M) represent terminal and middle nitrogen atoms of azide anion, respectively.}
\end{figure}

\par In order to explain the chemical bonding in NH$_4$N$_3$, we have calculated total and partial density of states (PDOS); see Fig. 3. The conduction band is mainly dominated by $2p$-states of nitrogen atoms (N(T), N(M), N of NH$_4$) and $1s$-states of hydrogen (H) atom. The top of valence band is completely dominated by $2p$-states of terminal nitrogen N(T) of azide ion, which implies that these states might be responsible for initiation of decomposition process in NH$_4$N$_3$. The relatively less contribution from $2p$-states of NH$_4$ ion, when compared to azide ion in the valence band region upto -4.5 eV from Fermi level, indicates a degree of non directional (ionic) bonding between ammonium (NH$^+_4$) and azide (N$^-_3$) ions. There is considerable overlap between $2p$-states of terminal N(T) and middle N(M) nitrogens of azide ion as well as $2p$-states of nitrogen from NH$_4$ and $s$-states of nitrogen atoms of azide ion in the energy range from -7.5 to -5 eV. Similarly, in the energy range from -10.5 to -7.5 eV, $2p$-states of nitrogen atom from NH$_4$ overlap with $1s$-states of the hydrogen atom. This does suggest that there is hybridization between N(T)-N(M) of azide ion and H-N of NH$_4$, implying directional (covalent) bonding in N-N (nitrogen atoms of azide ion) and N-H bonds. The Pauling scale has been used as measure of the attraction ability of an atom for electrons in a covalent bond.\cite{pauling} The difference in Pauling electronegativities of nitrogen (3.0) and hydrogen (2.1) is used to predict the nature of the N-H bond in ammonium ion. The electronegativity difference is 0.9, which points to a polar covalent bond character of N-H bond. In ammonium ion, the unit positive charge of the complex might be residing more on nitrogen atom, as a consequence of polar covalent character of N-H bond. This can be clearly understand from the electronic charge density contours along (200), (101), and (010) planes of NH$_4$N$_3$ crystal as shown in Fig. 4. The charge density along N(T)-N(M) is pronounced due to the hybridization between terminal and middle nitrogen of azide anion, suggesting directional bonding in azide ion. However, there is very less charge sharing between NH$^+_4$ and N$^-_3$ (see Fig. 3) ions indicates that ionic bonding predominant in NH$_4$N$_3$. The nature of bonding in NH$_4$N$_3$ is interpreted through bulk modulus, Colton \emph {et al} determined ionic character (in \%) for the AMAs and HMAs and found that AMAs are ionic solids while the HMAs are covalent solids.\cite{colton} Also, the bulk moduli of AMAs are lower when compared to HMAs (see section A), which reveals that the covalent azides have higher bulk modulus than ionic azides due to strong directional bonding. The bulk modulus of NH$_4$N$_3$ is found to be in middle of the AMAs and HMAs. Therefore, we confirmed from the calculated electronic band structure, DOS, charge density distributions, and bulk modulus, thus, NH$_4$N$_3$ exhibits predominantly ionic bonding along with partial covalent nature from the N-H and N-N bonds of NH$^+_4$ and N$^-_3$ ions, respectively.  

\begin{figure}[h]
\centering
\includegraphics[height = 4.5in, width=5.5in]{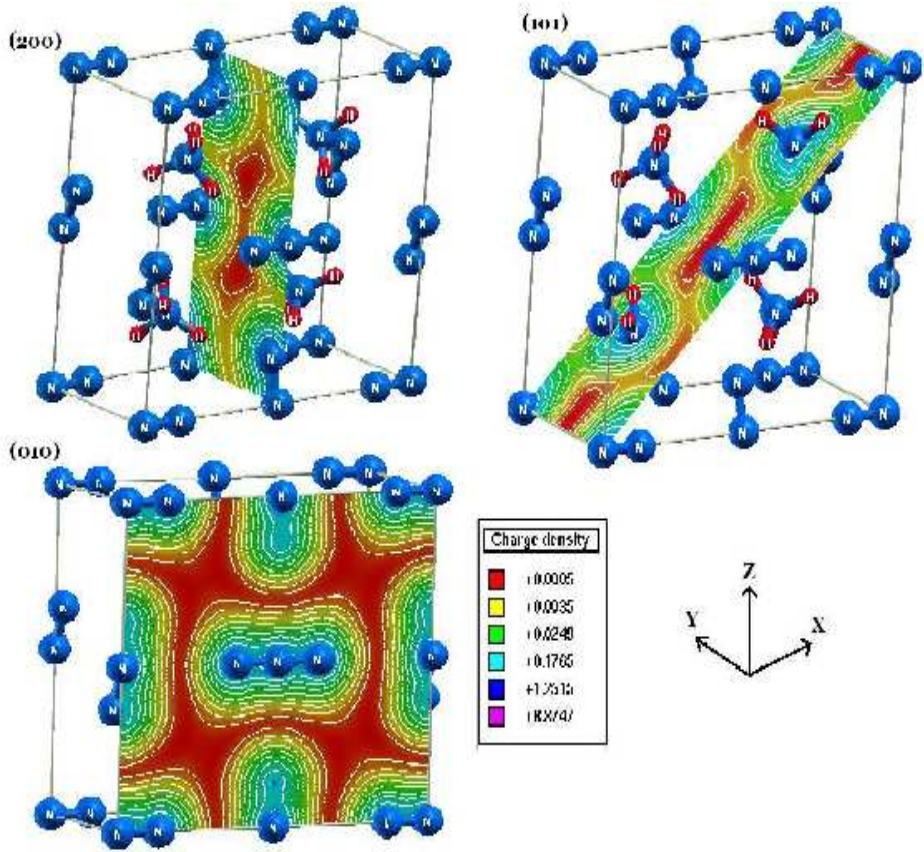}
\caption{ (Color online) Calculated electronic charge densities of NH$_4$N$_3$ in (200), (101), and (010) planes.}
\end{figure}

\begin{figure}[h]
\centering
\includegraphics[height = 4.5in, width=5.5in]{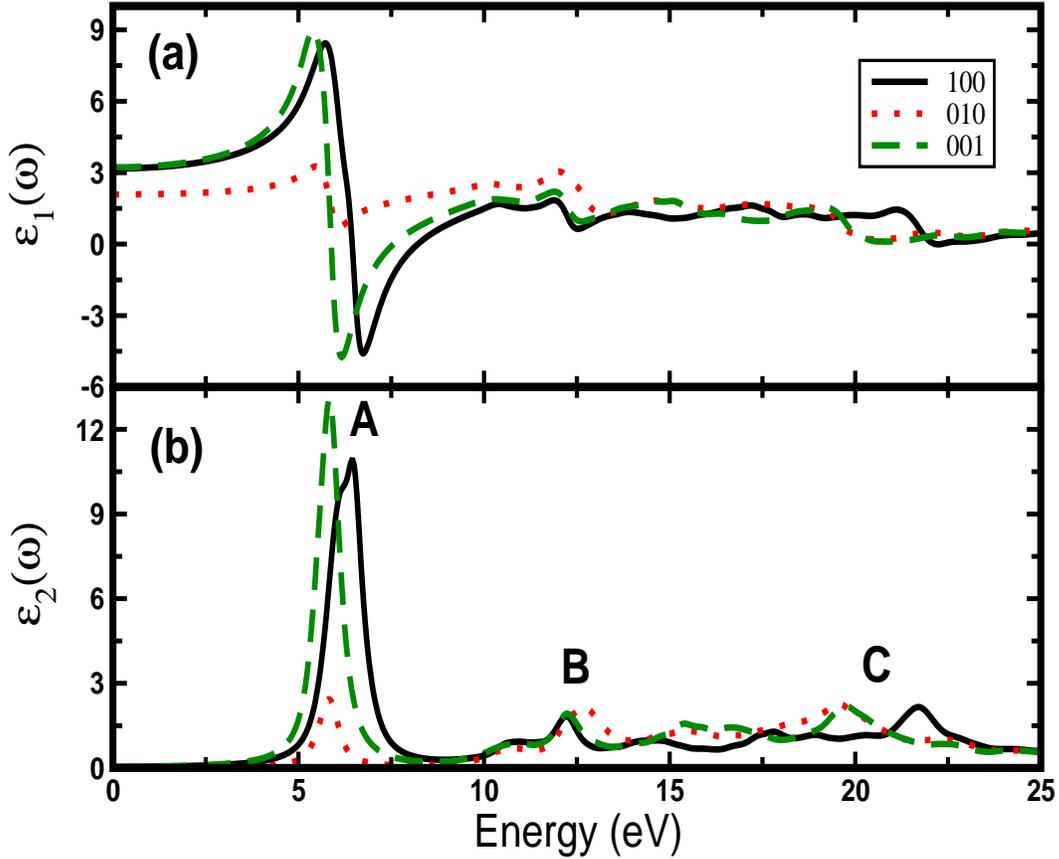}  \hspace{0.2in}
\caption{ (Color online) The real $\epsilon_1(\omega)$ and imaginary $\epsilon_2(\omega)$ parts of dielectric function of NH$_4$N$_3$ along three crystallographic axes as a function of photon energy at ambient conditions. Black solid line (along [100]), red dotted line (along [010]) and green thick dotted line (along [001]) .}
\end{figure}

\subsection{Optical properties}
The linear response of a system to electromagnetic radiation can be described by means of the dielectric function. In general, there are two contributions to dielectric function, namely; intra- and interband transitions. The intraband transitions occur only in metals. Further, the interband transitions are classified into direct and indirect transitions. The indirect interband transitions arise from scattering of phonons, which are neglected in our calculations because their contribution is negligible to the dielectric function when compared to direct interband transitions. The contribution of direct interband transitions to the imaginary $\epsilon_2(\omega)$ part of the dielectric function in the random phase approximation\cite{ehrenreich} without local field effects can be calculated by summing all the possible transitions between the occupied and unoccupied states for a set of k-vectors over the Brillouin zone. The real $\epsilon_1(\omega)$ part of dielectric function can be derived from the imaginary $\epsilon_2(\omega)$ part of dielectric function by using Kramers-Kronig relations.\cite{fox} 

\par The linear optical properties of NH$_4$N$_3$ have been calculated using TB-mBJ electronic structure with denser k-mesh of 735 k-points in the IBZ. NH$_4$N$_3$ crystallizes in the orthorhombic $Pmna$ space group; this symmetry has three independent components of dielectric function. Hence, the real and imaginary part of dielectric function are determined as a function of photon energy along three crystallographic ([100], [010], and [001]) directions as shown in Fig. 5. The imaginary part of dielectric function $\epsilon_2(\omega)$ has three prominent peaks due to interband transitions between valence band maximum (VBM) and conduction band minimum (CBM) along three crystallographic axes. The peaks in $\epsilon_2(\omega)$ (see Fig. 5(b)) are divided into three energy regions, named as A, 5-7.5 eV; B, 11.5-13.5 eV; and C, 17.5-23.5 eV. The peaks in the region A are due to interband transitions between $2p$ states of terminal nitrogen N(T) of azide ion to s-states of H/N of NH$_4$, while the peaks in region B as result of transitions between N-$2p$ of NH$_4$ to H-$1s$ states. Finally, the peaks in region C originate from H-$1s$ to N-$2p$ states of NH$_4$ and vice versa. It should be noted that most of the optical transitions are mainly from $2p$ (N) $\rightarrow$ $1s$ (H) states. The static real part of dielectric function $\epsilon_1(\omega)$ along three crystallographic directions is found to be 3.18 (along [100]), 2.09 (along [010]), and 3.22 (along [001]). Further, we have derived the four important optical constants absorption spectra $\alpha(\omega)$, refractive index n($\omega$), reflectivity R($\omega$), and loss function L($\omega$) of NH$_4$N$_3$ as a function of photon energy from the calculated real and imaginary parts of dielectric function using the formulations given in the ref. \onlinecite{fox}. They have been displayed in Fig. 6. As shown from Fig. 6(a), the absorption starts from 5.08 eV, which is a fundamental energy gap between VBM and CBM known as fundamental absorption edge for NH$_4$N$_3$. The first absorption peaks along three crystallographic directions [100], [010], and [001] are at 6.64, 5.94 and 6.03 eV and the corresponding absorption coefficients 1.7 x 10$^8$m$^{-1}$,  4.7 x 10$^7$m$^{-1}$, and 1.6 x 10$^8$m$^{-1}$, respectively. This result reveals that NH$_4$N$_3$ decomposes under the action of UV light. The absorption takes place from 5.08 (fundamental absorption edge) to 25 eV and then it decreases in the high energy region because the crystal becomes transparent above 25 eV. 

\begin{figure}[h]
\centering
\includegraphics[height = 5.0in, width=6.0in]{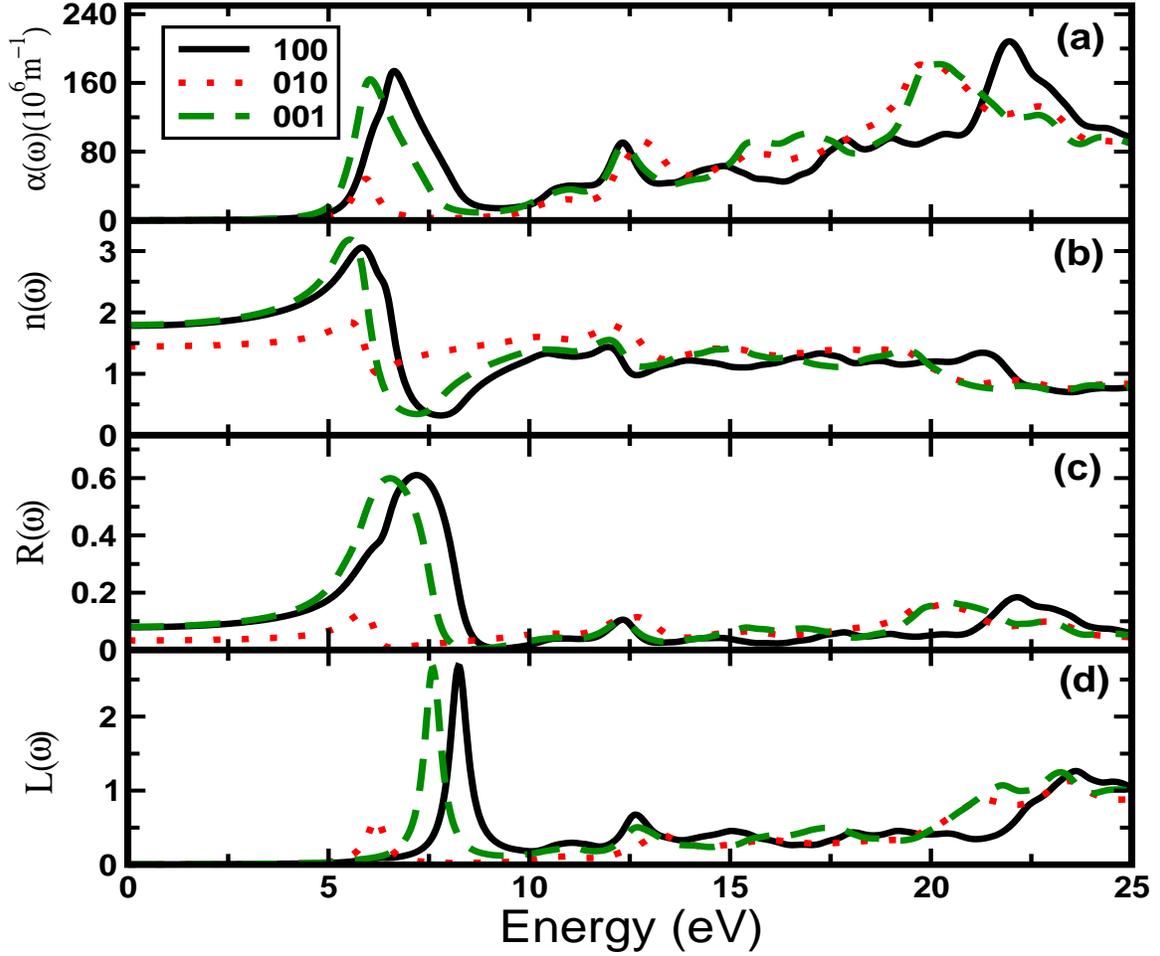}
\caption{ (Color online) The calculated absorption spectra $\alpha(\omega)$, refractive index n($\omega$), reflectivity R($\omega$) and loss function L($\omega$) of NH$_4$N$_3$ as a function of photon energy at ambient conditions. Black solid line (along [100]), red dotted line (along [010]) and green thick dotted line (along [001]).}
\end{figure} 
The calculated static refractive index of NH$_4$N$_3$ along the crystallographic axes is 1.78 (along [100]), 1.45 (along [010]), and 1.80 (along [001]). The distinct static refractive index values in each crystallographic direction indicates that NH$_4$N$_3$ is a optically anisotropic crystal (see Fig. 6(b)). The electron energy-loss function L($\omega$) describes the energy loss of a fast electron traversing in a material. The peaks in the L($\omega$) spectra represent the characteristics associated with the plasma resonance, and the corresponding frequency is the so-called plasma frequency, above which the material is a dielectric ($\epsilon_1(\omega)\textgreater$0) and below which the material behaves like a metallic compound ($\epsilon_1(\omega) \textless$0). As illustrated in Fig. 6(d), the prominent peaks in L($\omega$) correspond to trailing edges in the reflection spectra (see Fig. 6(c)) are observed at 8.2 eV (along [100]), 6.1eV (along [010]), and 7.6 eV (along [001]) directions and about 23.5 eV in all the three directions corresponding to abrupt reduction in the reflection spectra R($\omega$).

\subsection{Detonation properties}
For an explosive, detonation velocity (D) and detonation pressure (P) are the most important factors to evaluate its performance. The Kamlet-Jacobs equations\cite{kamlet, ablard} based on the reported density and heat of formation (HOF) were chosen for the prediction of detonation performance. Values for D (in km/s) and P (in GPa) were calculated according to the following equations:
\begin{center}
D = 1.01(NM$^{0.5}$Q$^{0.5}$)$^{0.5}$(1 + 1.30$\rho_0$) and P = 1.55$\rho^2_0$NM$^{0.5}$Q$^{0.5}$ \\
\end{center}
In the above equations, N is moles of gaseous detonation products per gram of explosives, M is average molecular weights of gaseous products, Q is chemical energy of detonation (kJ/mol) defined as the difference of the HOFs between products and reactants, and $\rho_0$ is the density of explosive (g/cc). We calculated the detonation velocity and pressure using calculated density (1.357 g/cc) and reported HOF (112.8 kJ/mol)\cite{williams,gray} and the corresponding values are found to be 6.45 km/s and 15.16 GPa, respectively. 

\section{CONCLUSIONS}
In the present study, first principles calculations were performed to investigate the structural, elastic, electronic and optical properties of NH$_4$N$_3$. The standard DFT functionals such as LDA, GGA are unable to account for the non-bonded dispersive forces in NH$_4$N$_3$ (see Table I). The calculated ground state properties using TS scheme are in excellent agreement with the experiment. Hence, we have used this scheme to calculate the elastic constants of NH$_4$N$_3$ and it is found to be mechanically stable. The observed ordering of the elastic constants is C$_{11}$ $\approx$ C$_{33}$ $\textgreater$ C$_{22}$, implying that NH$_4$N$_3$ is sensitive along b-axis due to weak intermolecular interactions between the atoms along this axis. We have derived the polycrystalline properties such as bulk modulus, shear modulus, density and Debye temperature from the calculated elastic constants. Finally, our PW-PP calculations confirm that the semiempirical schemes (TS, G06) can successfully treat the van der Waals interactions in NH$_4$N$_3$. The TB-mBJ electronic structure shows that NH$_4$N$_3$ is direct band gap insulator with an energy gap of 5.08 eV. The calculated PDOS and charge density contours show that strong ionic and weak covalent bonding nature in NH$_4$N$_3$. This is also confirmed by the obtained equilibrium bulk modulus of NH$_4$N$_3$ by comparing with bulk moduli of various inorganic azides. The calculated static refractive index and dielectric constants along the three different crystallographic axes indicates a strong anisotropy in NH$_4$N$_3$. Also, the absorption spectra reveal that NH$_4$N$_3$ is sensitive to UV light. Finally, the detonation velocity and detonation pressure are calculated using Kamlet-Jacobs equations from experimental heat of formation and calculated density. NH$_4$N$_3$ shows moderate performance, and the predicted detonation velocity and detonation pressure are 6.45 km/s and 15.16 GPa, respectively.   

\section{Acknowledgments}
NYK and VDG would like to thank DRDO through ACRHEM for financial support, and the CMSD, University of Hyderabad, for providing computational facilities. We thank Professor C. S. Sunandana, School of Physics, University of Hyderabad, for critical reading of the manuscript. We are also greatful to referees for their valuable suggestions and comments to improve the quality of the manuscript. \\
$^*$\emph{Author for Correspondence, E-mail: gvsp@uohyd.ernet.in}

\clearpage 
\bf Table of content (TOC) graphic \\
\begin{figure}[h]
\centering

\includegraphics[height = 5.0cm, width=5.0cm]{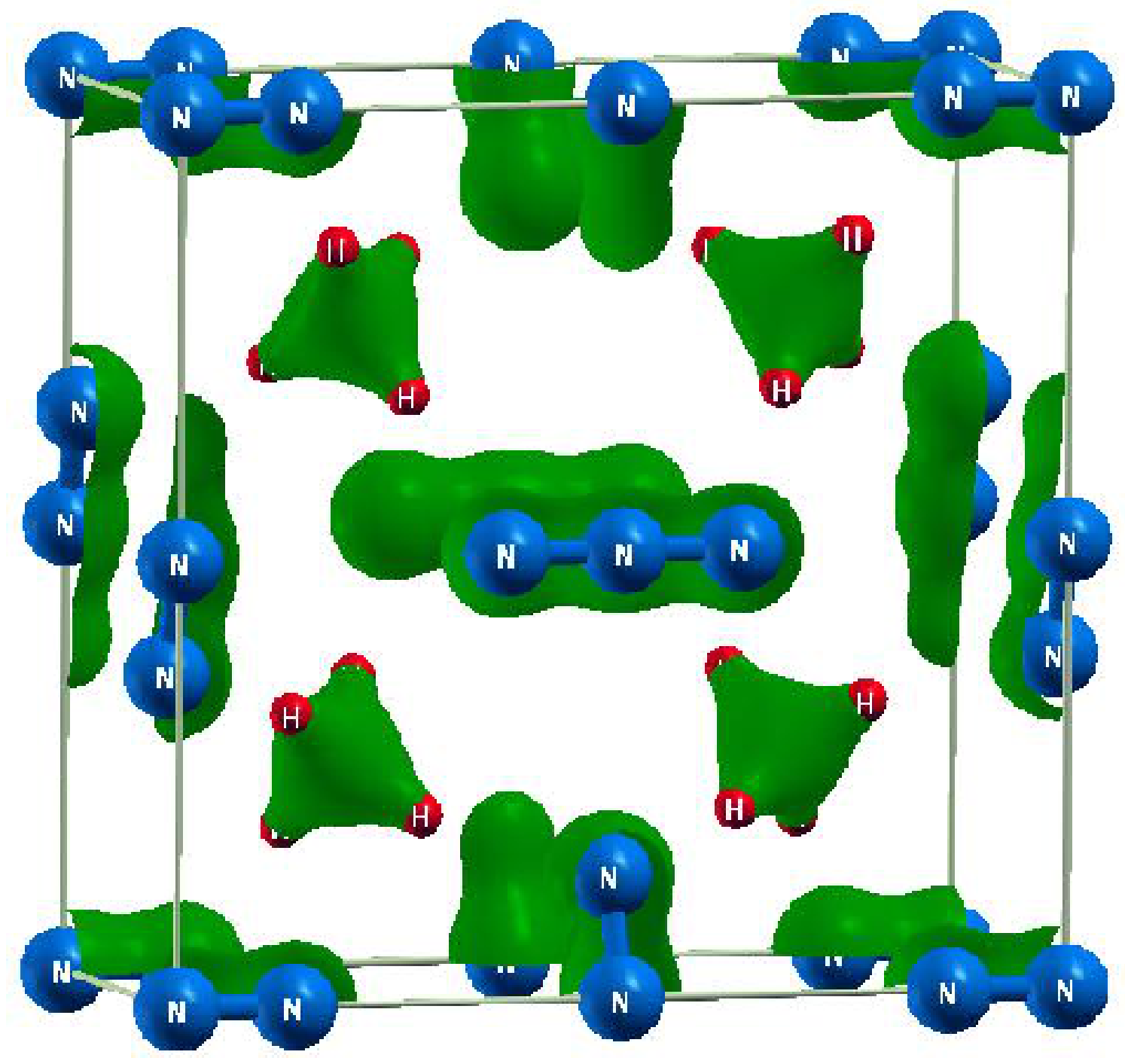}
\end{figure}


\clearpage
\begin{thebibliography}{plain}

\bibitem{fair}
Fair, H. D.; Walker, R. F. \emph{Energetic Materials}; Plenum Press: Newyork, 1977, Vol.1.

\bibitem{yoffe}
Bowden, F. P.; Yoffe, A. D. \emph{Fast Reactions in Solids}; Butterworth: London, U.K., 1958.

\bibitem{hemley} 
Eremets, M. I.; Hemley, R. J.; Mao, H. K.; Gregoryanz, E. \emph{Nature} {\bf 2001}, \emph{411}, 170-174.

\bibitem{eremets} 
Eremets, M. I.; Gavriliuk, A. G.; Trojan, I. A.; Dzivenko, D. A.; Boehler, R. \emph{Nat. Matter.} {\bf 2004}, \emph{3}, 558-563.

\bibitem{trojen}
Eremets, M. I.; Popov, M. Y.; Trojan, I. A.; Denisov, V. N.; Boehler, R; Hemley, R. J. \emph{J. Chem. Phys.} {\bf 2004}, 10618-10623. 

\bibitem{medvedev}
Medvedev, S. A.; Eremets, M. I.; Evers, J; Klapotke, T. M.; Palasyuk, T.; Trojan, I. A. \emph{Chem. Phys.} {\bf 2011}, \emph{386}, 41-44.

\bibitem{palasyuk}
Medvedev, S. A.; Palasyuk, T.; Trojan, I. A.; Naumov, P. G.; Evers, J; Klapotke, T. M.; Eremets, M. I. \emph{Vib. Spectrosc.} {\bf 2012}, \emph{58}, 188-192.

\bibitem{prince}
Prince, E.; Choi, C. S. \emph{Acta Crystallogr. Sect. B} {\bf 1978}, \emph{34}, 2606-2608.

\bibitem{curtius}
Curtius, T. \emph{Ber. Deutsch. Chem. Gesellschaft} {\bf 1890}, \emph{23}, 3023-3033.

\bibitem{eslami}
Eslami, A.; Hosseini, S. G.; Shariaty, S. H. M. \emph{Powder Technol.} {\bf 2011}, \emph{208}, 137-143.

\bibitem{ng}
Ng, W. L.; Field, J. E. \emph{Thermochim. Acta} {\bf 1985}, \emph{84}, 133-140. 

\bibitem{evers}
Evers, J.; Gobel, M.; Krumm, B.; Martin, F.; Medvedyev, S.; Oehlinger, G.; Steemann, F. X.; Troyan, I.; Klapotke, T. M.; Eremets M. I. \emph{J. Am. Chem. Soc.} {\bf 2011}, \emph{133}, 12100-12105.     

\bibitem{yakovleva}
Yakovleva, G. S.; Kurgangalina R. K.; Stestik, L. N. Fiz. \emph{Goreniya Vzryva} {\bf 1977}, \emph{13}, 473-475.

\bibitem{masaaki}
Masaaki, S.; Takehiko, S.; Tadamasa, H.; Ikuo, H. \emph{Japan Kokai} {\bf 1974}, \emph{74}, 08475.

\bibitem{maycock}
Maycock, J. N.; Pai Verneker, V. R. \emph{J. Spacecr. Rockets} {\bf1969}, \emph{6}, 336-337.

\bibitem{sarner}
Sarner, S. R. \emph{Propellant Chemistry}; Van Nostrand Reinhold: New York, 1966.

\bibitem{hu}
Hu, A.; Zhang, F. \emph{J. Phys.: Condens. Matter} {\bf 2011}, \emph{23}, 022203.

\bibitem{kurgangalina}
Yakovleva, G. S.; Kurgangalina R. K.; Stesik, L. N. \emph{Combust. Explos. Shock} {\bf 1997}, \emph{13}, 405-407.

\bibitem{finch}
Finch, A.; Gardner, P. J.; Head, A. J.; Xiaoping, W. \emph{J. Chem. Thermodyn.} {\bf 1990}, \emph{22}, 301-305. 

\bibitem{Payne} 
Payne, M. C.; Teter, M. P.; Allen, D. C.; Arias, T. A.; Joannopoulos, J. D. \emph{Rev. Mod. Phys.} \textbf{1992}, \emph{64}, 1045-1097.

\bibitem{Milman}
Milman, V.; Winkler, B.; White, J. A.; Packard, C. J.; Payne, M. C.; Akhmatskaya, E. V.; Nobes, R. H. \emph{Int. J. Quantum Chem.} \textbf{2000}, \emph{77}, 895-910.

\bibitem{Segall}
Segall, M. D.; lidan, P. L. D.; Probert, M. J.; Pickard, C. J.; Hasnip, P. J.; Clark, S. J.; Payne, M. C. \emph{J. Phys.: Condens. Matter} \textbf{2002}, \emph{14}, 271-280.

\bibitem{Hohenberg} 
Hohenberg, P.; Kohn, W. \emph{Phys. Rev. B} {\bf 1964}, \emph{136}, 384-389.

\bibitem{Kohn} 
Kohn, W.; Sham, L. J. \emph{Phys. Rev. A} {\bf 1965}, \emph{140}, 1133-1138.

\bibitem{Vanderbilt} 
Vanderbilt, D. \emph{Phys. Rev. B} \textbf{1990}, \emph{41}, 7892-7895.

\bibitem{Ceperley} 
Ceperley, D. M.; Alder, B. J. \emph{Phys. Rev. Lett.} \textbf{1980}, \emph{45}, 566-569.

\bibitem{Perdew} 
Perdew, J. P.; Zunger, A. \emph{Phys. Rev. B} \textbf{1981}, \emph{23}, 5048-5079.

\bibitem{Burke} 
Perdew, J. P.; Burke, S.; Ernzerhof, M. \emph{Phys. Rev. Lett.} {\bf 1996}, 77, 3865-3868.

\bibitem{Santra}
Santra, B.; Klimes, J.; Alfe, D.; Tktchenko, A.; Slater, B.; Michaelides, A.; Car, R.; Scheffler, M. \emph{Phys. Rev. Lett.} \textbf{2011}, \emph{107}, 185701.

\bibitem{Lu}
Lu, D.; Li, Y.; Rocca, D.; Galli, G. \emph{Phys. Rev. Lett.} \textbf{2009}, \emph{102}, 206411.

\bibitem{Dion}
Dion, M.; Rydberg, H.; Schroder, E.; Langreth, D. C.; Lundqvist, B. I. \emph{Phys. Rev. Lett.} \textbf{2004}, \emph{92}, 246401.

\bibitem{Grimme}
Grimme, S. \emph{J. Comput. Chem.} \textbf{2006} \emph{27}, 1787-1799.

\bibitem{Tkatchenko}
Tkatchenko, A.; Scheffler, M. \emph{Phys. Rev. Lett.} \textbf{2009} \emph{102}, 073005.

\bibitem{Almlof}
Fischer, T. H.; Almlof, J. \emph{J. Phys. Chem.} {\bf 1992}, \emph{96}, 9768-9774.

\bibitem{Monkhorst} 
Monkhorst, H. J.; Pack, J. D. \emph{Phys. Rev. B} \textbf{1976}, \emph{13}, 5188-5192.

\bibitem{Nieminen}
Nieminen, R. M. \emph{Modelling Simul. Mater. Sci. Eng.} \textbf{2009}, \emph{17}, 084001.

\bibitem{Sham}
Sham, L. J.; Schluter, M. \emph{Phys. Rev. Lett.} {\bf 1983}, \emph{51}, 1888.

\bibitem{peter} 
Tran, F.; Blaha, P. \emph{Phys. Rev. Lett.} {\bf 2009}, \emph{102}, 226401.

\bibitem{blaha} 
Blaha, P.; Schwarz, K.; Madsen, G. K. H.; Kvasnicka, D.; Luitz, J. \emph{WIEN2K}, an Augmented Plane Wave + Local Orbitals Program for Calculating Crystal Properties, Karlheinz Schwarz, Technische. Universitat: Wien, Austria. 2001.

\bibitem{amorim}
De Amorim, H. S.; do Amaral Jr., M. R.; Pattison, P.; Ludka, I. P.; Mendes, J. C. \emph{J. Mex. Chem. Soc.} \textbf{2002}, \emph{46}, 313-319.

\bibitem{murnaghan}
Murnaghan, F. D. Proc. Natl. Acad. Sci. U.S.A. \textbf{1944}, \emph{30}, 244-247. 

\bibitem{medevdev}
Medevdev, S. A.; Trojan, I. A.; Eremets, M. I.; Palasyuk, T.; Klapotke, T. M.; Evers, J. \emph{J. Phys.: Condens. Matter} \textbf{2009}, \emph{21}, 195404.

\bibitem{weir}
Weir, C. E.; Block, S.; Piermarini, G. J. \emph{J. Chem. Phys.} \textbf{1970}, \emph{53}, 4265-4269.

\bibitem{cheng}
Ji, C.; Zhang, F.; Hou, D.; Zhu, H.; Wu, J.; Chyu, M. C.; Levitas, V. I.; Ma Y. \emph{J. Phys. Chem. Solids} \textbf{2011}, \emph{72}, 736-739.

\bibitem{dongbin}
Hou, D.; Zhang, F.; Ji, C.; Hannon, T.; Zhu, H.; Wu, J.; Ma, Y. \emph{Phys. Rev. B} \textbf{2011}, \emph{84}, 064127.

\bibitem{hou}
Hou, D.; Zhang, F.; Ji, C.; Hannon, T.; Zhu, H.; Wu, J.; Levitas, V. I.; Ma, Y. \emph{J. Appl. Phys.} \textbf {2011}, \emph{110}, 023524.

\bibitem{Mehl} 
Mehl, M. J.; Osburn, J. E.; Papaconstantopoulus, D. A.; Klein, B. M. \emph{Phys. Rev.B} {\bf 1990}, \emph{41}, 10311-10323.

\bibitem{Born} 
Born, M.; Huang, K. \emph{Dynamical Theory of Crystal Lattices}; Oxford University Press: Oxford, U.K., 1998.

\bibitem{haycraft}
Haycraft, J. J.; Stevens, L. L.; Eckhardt, C. J. \emph{J. Chem. Phys.} {\bf 2006}, \emph{124}, 024712.

\bibitem{stevens}
Stevens, L. L.; Eckhardt, C. J. \emph{J. Chem. Phys.} {\bf 2005}, \emph{122}, 174701.

\bibitem{eckhardt}
Eckhardt, C. J. \emph{J. Chem. Phys.} {\bf 2009}, \emph{131}, 214501.

\bibitem{reuss} 
Reuss, A; Angew, Z. \emph{Math. phys.} {\bf 1929}, \emph{9}, 49.

\bibitem{ravindran}
Ravindran, P.; Fast, L.; Korzhavyi, P. A.; Johansson, B. \emph{J. Appl. Phys.} \textbf{1998}, \emph{84}, 4891-4904.

\bibitem{pugh} 
Pugh, S. F. \emph{Philos. Mag.} \textbf{1954}, \emph{45}, 823-843.

\bibitem{pettifor}
Pettifor, D. G. \emph{Mater. Sci. Technol.} \textbf{1992}, \emph{8}, 345-349.

\bibitem{anderson}
Anderson, O. L. \emph{J. Phys. Chem. Solids} \textbf{1963}, \emph{24}, 909-917.

\bibitem{colton}
Colton, R. J.; Rabalais, J. W. \emph{J. Chem. Phys.} \textbf{1976}, \emph{64}, 3481-3486.

\bibitem{zhu1}
Zhu, W. H.; Xiao, H. M. \emph{Struct. Chem.} {\bf 2010}, \emph{21}, 847-854.

\bibitem{hedin1}
Hedin, L. \emph{Phys. Rev.} \textbf{1965}, \emph{139}, A796-A823.

\bibitem{hedin2}
Hedin, L.; Lundquist, S. \emph{Solid State Physics}; Academic Press: NewYork, 1969; Vol. 23, p 1.

\bibitem{koller1}
Koller, D.; Tran, F.; Blaha, P. \emph{Phys. Rev. B} \textbf{2012}, \emph{85}, 155109.

\bibitem{koller2}
Koller, D.; Tran, F.; Blaha, P. \emph{Phys. Rev. B} \textbf{2011}, \emph{83}, 195134.

\bibitem{dixit}
Dixit, H.; Saniz, R.; Cottenier, S.; Lamoen, D.; Partoens, B. \emph{J. Phys.: Condens. Matter} \textbf{2012}, \emph{24}, 205503.

\bibitem{singh} 
Singh, D. J. \emph{Phys. Rev. B} \textbf{2010}, \emph{82}, 205102. 

\bibitem{zhu2}
Zhu, W. H.; Xiao, H. M. \emph{J. Comput. Chem.} {\bf 2008}, \emph{29}, 176-184.

\bibitem{zhu3}
Zhu, W. H.; Xiao, J. J.; Ji, G. F.; Zhao, F.; Xiao, H. M. \emph{J. Phys. Chem. B} {\bf 2007}, \emph{111}, 12715.

\bibitem{zhu4}
Xu, X. J.; Zhu, W. H.; Xiao, H. M. \emph{J. Phys. Chem. B} {\bf 2007}, \emph{111}, 2090-2097.

\bibitem{kuklja1}
Kuklja, M. M.; Stefanovich, E. V.; Kunz, A. B. \emph{J. Chem. Phys.} {\bf 2000}, \emph{112}, 3417-3423.

\bibitem{kuklja2}
Kuklja, M. M.; Kunz, A. B. \emph{J. Appl. Phys.} \emph{2000}, \emph{87}, 2215-2218.

\bibitem{pauling}
Pauling, L. \emph{The Nature of the Chemical Bond}, 3rd ed.; Cornell University Press: Ithaca, NY, 1960.

\bibitem{ehrenreich}
Ehrenreich, H.; Cohen, M. H. \emph{Phys. Rev.} \textbf{1959}, \emph{115}, 786-790.

\bibitem{fox}
Fox, M. \emph{Optical Properties of Solids}; Oxford University press: Newyork, 2001.

\bibitem{kamlet} 
Kamlet, M. J.; Jacobs, S. J. \emph{J. Chem. Phys.} {\bf 1968}, \emph{48}, 23-35.

\bibitem{ablard} 
Kamlet, M. J.; Ablard, J. E. \emph{J. Chem. Phys.} {\bf 1968}, \emph{48}, 36-42.

\bibitem{williams}
Williams, L. O. US Patent 5,081,930, 1992.  

\bibitem{gray}
Gray, P.; Waddington, T. C. \emph{Proc. R. Soc. Lond., A} \textbf{1956}, \emph{235}, 481-495.

\end{thebibliography}
\end{document}